\documentclass[pra,showpacs,nofootinbib]{revtex4}
\usepackage{hyperref}
\usepackage{amsmath}
\usepackage{amssymb}
\usepackage{amsthm}
\usepackage{graphics}
\usepackage{graphicx}
\usepackage{color}
\usepackage[usenames,dvipsnames,svgnames,table]{xcolor}

\long\def\ca#1\cb{}

\def\Tr#1{\textrm{Tr}(#1)}

\def\DC{{\cal D}}

\def\HC{{\cal H}}

\def\KC{{\cal K}}

\def\SC{{\cal S}}

\def\endproof{{\hspace{\stretch{1}}$\blacksquare$}}

\newtheorem{thm1}{Theorem}
\newtheorem{thm2}[thm1]{Theorem}
\newtheorem{thm3}[thm1]{Theorem}
\newtheorem{thm4}[thm1]{Theorem}
\newtheorem{lem1}[thm1]{Lemma}

\begin{document}
\title{Extended necessary condition for local operations and classical communication: Tight bound for all measurements}
\author{Scott M. Cohen}
\email{cohensm52@gmail.com}
\affiliation{Department of Physics, Portland State University, Portland OR 97201}

\begin{abstract}
We give a necessary condition that a separable measurement can be implemented by local quantum operations and classical communication (LOCC) in any finite number of rounds of communication, generalizing and strengthening a result obtained previously. That earlier result involved a bound that is tight when the number of measurement operators defining the measurement is relatively small. The present results generalize that bound to one that is tight for any finite number of measurement operators, and we also provide an extension which holds when that number is infinite. We apply these results to the famous example on a $3\times3$ system known as ``domino states", which were the first demonstration of nonlocality without entanglement. Our new necessary condition provides an additional way of showing that these states cannot be perfectly distinguished by (finite-round) LOCC. It directly shows that this conclusion also holds for their cousins, the rotated domino states. This illustrates the usefulness of the present results, since our earlier necessary condition, which these results generalize, is not strong enough to reach a conclusion about the domino states.
\end{abstract}

\date{\today}

\maketitle
\section{Introduction}
In a recent publication \cite{myExtViolate1}, we proved a necessary condition such that a quantum measurement can be implemented by local operations on subsystems and classical communication between parties (LOCC) in any finite number of rounds of communication.\footnote{We define each round of communication as consisting of one party broadcasting the result of her measurement to all the other parties.} We also demonstrated that there exist examples of separable measurements for which the condition is extensively violated, this violation growing without limit as the number of parties increases. A class of the examples given in \cite{myExtViolate1} was later shown \cite{myUSD} to be applicable to the optimal unambiguous discrimination of certain sets of quantum states, and includes an infinite number of cases for each number of parties where each case is such that separable measurements are strictly better than finite-round LOCC. We also discussed in \cite{myExtViolate1} why we believe that these results apply to all LOCC, including those using an infinite number of rounds, but to date a proof remains elusive.

Each quantum measurement involves a set of operators $K_j$ where, for the $j$th outcome of the measurement, the state of the measured system changes as $\rho\rightarrow K_j\rho K_j^\dag/p_j$,  with $p_j=\Tr{\rho\KC_j}$, where the associated `POVM element' is defined as $\KC_j:=K_j^\dag K_j$. A measurement on $P$ parties is separable \cite{Rains} if and only if each $K_j$ is a product operator, in which case the POVM elements are also product, $\KC_j=\KC_j^{(1)}\otimes\ldots\otimes\KC_j^{(P)}$. It is well-known that every LOCC measurement is separable, but there exist separable measurements that are not LOCC \cite{Bennett9,IBM_CMP,IBM_PRL,NisetCerf}. In an effort to better understand the difference between separable measurements and LOCC \cite{ChitambarDuan,StahlkeU,YingSEP,Koashi,ChildsLeung,ChitCuiLoPRA,ChitCuiLoPRL,Gheorghiu1}, we have undertaken a series of works \cite{mySEPvsLOCC,myMany,myExtViolate1,myUSD} aimed at finding conditions on the sets of POVM elements that could serve to distinguish between these two important classes of quantum measurements.

In \cite{mySEPvsLOCC}, we showed how to construct an LOCC protocol for a given bipartite separable measurement whenever such a protocol exists in a finite number of rounds. This result was generalized to any number of parties in \cite{myMany}. The approach in these papers involves looking for intersections of convex cones generated by subsets of the local operators $\KC_j^{(\alpha)}$ associated with the measurement under consideration, and the starting point is to consider subsets that each consist of a single operator. Each operator $\KC_j^{(\alpha)}$ generates a ray $\{\lambda\KC_j^{(\alpha)}\vert\lambda\ge0\}$, any collection of these rays generates a convex cone, and the extreme rays of these cones are those associated with operators in that collection which cannot be written as a positive linear combination of the others in the same collection. Clearly, if for each party $\alpha$, every $\KC_j^{(\alpha)}$ is extreme in the cone of the full set of these operators for a given measurement and no two $\KC_j^{(\alpha)}$ are proportional, then the starting point mentioned above will fail, as no two cones involving just a single operator will intersect. More generally, it appeared that (loosely speaking) too many extreme rays would make it difficult to find enough intersections to build a full LOCC protocol for the measurement. Motivated by this idea, we proved the following theorem in \cite{myExtViolate1}.
\begin{thm1} \label{thm1}\cite{myExtViolate1} For any finite-round LOCC protocol of $P$ parties implementing a separable measurement corresponding to the $N$ distinct POVM elements $\{\KC_j=\KC_j^{(1)}\otimes\ldots\otimes\KC_j^{(P)}\}_{j=1}^N$, it must be that
\begin{align}\label{eqn11}
\sum_{\alpha=1}^P e_\alpha\le2(N-1),
\end{align}
where $e_\alpha$ is the number of distinct extreme rays in the convex cone generated by operators $\{\KC_j^{(\alpha)}\}_{j=1}^N$, and the sum includes only those parties for which at least one of these local operators is not proportional to the identity. The upper bound in \eqref{eqn11} can be achieved with equality when $N\le2^P$.
\end{thm1}

The last line in this theorem, that the upper bound in \eqref{eqn11} is tight for $N\le2^P$, is rather restrictive. In general, $N$ will be larger, in many cases significantly so. For example, any measurement aimed at discriminating a full basis of states will consist of $N=d_1d_2\ldots d_P$ operators ($d_\alpha$ is the dimension of the Hilbert space $\HC_\alpha$ for subsystem $\alpha$), which exceeds $2^P$ unless all subsystems are qubits, the smallest nontrivial system. Therefore, it would be of interest to have an upper bound that is tight for $N>2^P$, as well. At the time of writing of \cite{myExtViolate1}, we had been able to prove that for bipartite systems, $\sum_\alpha e_\alpha\le3N/2$, which is a tight bound whenever $N\ge2^P=4$. We had suspected that an upper bound of $2N(1-2^{-P})$ might be valid for all $P$, but our method of proof for bipartite systems could not be generalized to more than two parties and at the time, we saw no way of approaching it differently. Recently, we realized that the techniques of \cite{myExtViolate1} could be used to prove this conjectured upper bound, which is tight for $N\ge2^P$ and any $P$. We present the results of this approach in Theorems~\ref{thm2}, \ref{thm3}, and \ref{thm4}, below. The latter two theorems apply to the case of infinite $N$.

We start by representing any LOCC measurement by a canonical LOCC tree, as defined in \cite{myExtViolate1}, a representation which is possible for any measurement implemented by LOCC. Each node is labeled by the POVM element corresponding to the cumulative action, to that point in the protocol, of the party for whom that node represents one outcome of a measurement by that party. If that party who measured is $\alpha$, we refer to that node as an $\alpha$-node. A canonical LOCC tree is then one where every nonleaf node has exactly two child nodes, and for any given node, the pair of POVM elements labeling its two child nodes are not proportional to each other. Given this structure, these are full binary trees, so if they have $N$ leaf nodes, they will also have a total of $2N-1$ nodes in all \cite{FullBinaryTreeThm}. 

In the following section, we state and prove the finite-$N$ result, Theorem~\ref{thm2}. Its infinite-$N$ counterparts, Theorems~\ref{thm3} and \ref{thm4}, are also stated in this section, and their proof is given in the appendix. In Section~\ref{sec3}, we present physically motivated tasks for which Theorem~\ref{thm2} demonstrates directly that these tasks cannot be accomplished by finite-round LOCC. Finally, in Section~\ref{sec4}, we offer our conclusions. As with Theorem~\ref{thm1}, we also conjecture that these theorems need not be restricted to finite-round LOCC, but rather apply to infinite-round LOCC, as well.

\section{Main results}\label{sec2}
In \cite{myExtViolate1}, we showed how any canonical LOCC tree can be pruned down to a full binary tree that has one and only one leaf for each distinct $\KC_j$ in the corresponding separable measurement, but nonetheless still has at least one appearance of each of the $\KC_j^{(\alpha)}$ that is an extreme ray, for every party $\alpha$. In addition, the method of pruning the tree is such that descendants of a given node in the final tree were also descendants of that node in the original tree. For our present result, consider an arbitrary canonical LOCC protocol represented as a canonical LOCC tree. We want to count the number of extreme rays in this tree. The first step will be to prune the tree as described in \cite{myExtViolate1}, after which we rearrange the remaining nodes as follows: If a node is an extreme ray and at least one of its children is not extreme, swap positions of these two nodes, which moves the extreme node closer to the leafs of the tree. If both children are non-extreme, just choose either one to swap with its parent. Continue this process until no extreme node has a child that is not extreme, which means that any descendant of an extreme node must itself be extreme.

The resulting tree remains full binary and every extreme node lies in a subtree within which every node is extreme. Using integer index $s$, denote each maximal such subtree as $T_s$. If for given $s$, $T_s$ has $l_s$ leaf nodes, then as it is still full binary, it also has $2l_s-1$ nodes in total, each of which is extreme. Suppose there are $\SC$ of these subtrees. Then the total number of extreme nodes in this tree is equal to the total number of nodes in the collection of these subtrees, which is
\begin{align}\label{eqn21}
\sum_{s=1}^\SC(2l_s-1)=2\sum_{s=1}^\SC l_s-\SC\le2N-\SC,
\end{align}
where we have used the fact that the total number $N$ of leaf nodes is at least $\sum_sl_s$ (this sum can be strictly less than $N$ if there is one or more subtrees that have no extreme rays in them at all, since in this case the leafs in these subtrees are not counted in the sum). Even though no two leafs are the same $\KC_j$, it is still possible that some extreme rays are repeated at different nodes in this tree.\footnote{For example, it may be that $\KC_1^{(1)}=\KC_2^{(1)}$ is a (single) extreme ray, and these operators appear as two different leafs, one being the unique $\KC_1$ leaf, the other being the unique $\KC_2$ leaf. In this case, both these leaf nodes represent the same extreme ray in the first party's set of rays, and the number of extreme nodes in the tree is strictly greater than the number of extreme rays.} However, since every extreme ray appears at least once in the tree, the number of extreme rays is no greater than the number of extreme nodes, which is itself no greater than the quantity $2N-\SC$ on the right-hand side of \eqref{eqn21}. Therefore, we have that
\begin{align}\label{eqn22}
\sum_{\alpha=1}^Pe_\alpha\le2N-\SC.
\end{align}

If we can find the smallest possible value of $\SC$, this will give a good upper bound on the total number of distinct extreme rays. Recall from \cite{myExtViolate1} that the root node of the original tree is always present in the pruned tree and is not extreme. It should be clear that this node is still the root of the entire pruned and rearranged, final tree, implying $\SC\ge2$. In fact $\SC=2$ is possible, occurring when the root is the only non-extreme node in the pruned tree, and then no re-arrangement is necessary. In this case, $\sum_\alpha e_\alpha\le2(N-1)$, and we recover Theorem~\ref{thm1}.

It turns out, however, that $\SC=2$ is not always possible, depending on how many parties there are involved in the protocol. We will presently show that the number of leaf nodes in any one of these subtrees cannot exceed $2^{P-1}$. Now, $\SC$ is minimized when each subtree has this maximum number of leaf nodes, which occurs for these full binary subtrees when every branch has the same maximum height (height is the number of edges between the root and the leaf). The height of these subtrees is limited by the fact that every node in each of them is extreme, along with the fact that extreme $\alpha$-nodes have no $\alpha$-node descendants. The latter point is true of the original tree, by Lemma~5 of \cite{myExtViolate1}, and as is pointed out there, this remains true for the pruned tree. It also applies to the final, rearranged tree, since our rearrangement, like the pruning, does not create new descendants of any extreme node, but rather only turns some of its descendants into non-descendants. Therefore, no branch in these subtrees can have more than one $\alpha$-node, for each of the $P$ parties $\alpha$, which directly implies there are no more than $P$ nodes along any branch within any one of these subtrees, whose height $h$ must therefore satisfy $h\le P-1$. It is well-known for a binary tree with $l$ leaves and height $h$ that $l\le2^h$ \cite{BenderWilliamson}, so we can conclude that the number of leaves in any one of these subtrees cannot exceed $2^{P-1}$. Since there are a total of $N$ leaves in the full collection of these subtrees, there must be at least $N/2^{P-1}$ subtrees in this collection. Hence,
\begin{align}\label{eqn23}
\SC\ge\left\lceil\frac{N}{2^{P-1}}\right\rceil,
\end{align}
where $\lceil x\rceil$ is the smallest integer not less than $x$, and from \eqref{eqn22} we have
\begin{align}\label{eqn24}
\sum_{\alpha=1}^Pe_\alpha\le2N-\left\lceil\frac{N}{2^{P-1}}\right\rceil,
\end{align}
Combining this with Theorem~\ref{thm1}, we have
\begin{thm2} \label{thm2}For any finite-round LOCC protocol of $P$ parties implementing a separable measurement corresponding to the $N$ distinct POVM elements $\{\KC_j=\KC_j^{(1)}\otimes\ldots\otimes\KC_j^{(P)}\}_{j=1}^N$, it must be that
\begin{align}\label{eqn25}
\sum_{\alpha=1}^P e_\alpha\le2N-\left\lceil2N\delta\right\rceil,
\end{align}
where $\delta=\max\left(N^{-1},2^{-P}\right)$, $e_\alpha$ is the number of distinct extreme rays in the convex cone generated by operators $\{\KC_j^{(\alpha)}\}_{j=1}^N$, and the sum includes only those parties for which at least one of these local operators is not proportional to the identity. The upper bound in \eqref{eqn25} can be achieved with equality for any finite $N$ and $P$.
\end{thm2}

We showed in \cite{myExtViolate1} how the upper bound can be achieved when $N\le2^P$. The discussion above indicates how it can be done for all finite $N$. First consider the special case that $N=2^{P-1}n$ with integer $n\ge3$. One party measures first with $n$ distinct outcomes. For each of her outcomes, each of the other $P-1$ parties measures once with a two-outcome measurement along every branch, conditioning their measurements on the previous parties' outcomes. As a result, descended from each of the $n$ outcomes of that initial measurement, there is a full binary subtree having $2^{P-1}$ leaf nodes and $2^P-1$ nodes. This gives a total of $N=2^{P-1}n$ leafs in the entire tree, which also has a total of $\left(2^P-1\right)n=2N(1-\delta)$ nodes, not counting the root of the tree. The parties can choose their measurements so that all of their local outcomes are distinct from each other, and so that each such outcome is extreme in the cone of its collection of local outcomes. Then every node in the tree is extreme apart from the root of the tree, and the bound is achieved with equality.

If the last measurement along a single branch of the preceding protocol is omitted, this removes two leaf nodes, but the node that was the parent of those two removed leafs becomes a new leaf, so $N$ is decreased by one to $N=2^{P-1}n-1$. At the same time, the total number of nodes is decreased by two, as is the total number of extreme rays. Now, $\left\lceil2N\delta\right\rceil=\left\lceil2N/2^P\right\rceil$ doesn't change when $N$ decreases by one, so the upper bound in \eqref{eqn25} also decreases by two, and is again achieved with equality. This process can be continued sequentially, at each step omitting a single measurement in the same chosen subtree. The quantity $\left\lceil2N\delta\right\rceil$ remains unchanged as $N$ decreases by one and the number of extreme rays decreases by two, with the upper bound always being achieved with equality, until there is only one node left in that subtree. When that subtree's last node is removed, $N$ has decreased by $2^{P-1}$ in all, which is the point at which $\left\lceil2N\delta\right\rceil$ decreases by one. This last removal decreases $N$ by one, the number of extreme rays also by one, and the upper bound by one, so the upper bound is again achieved with equality. At this point we are effectively back where we started but with one less outcome in the first party's initial measurement, so start again omitting measurements in another subtree. By continuing this process even into the last remaining subtree, we see that the bound is tight for any finite $N$.

Let us now turn to the case of a separable measurement having an infinite number of distinct POVM elements. Begin by choosing an ordering of these POVM elements. Let $e_{\alpha N}$ be the number of distinct extreme rays for party $\alpha$ in the first $N$ of those POVM elements. Define the density of extreme rays as
\begin{align}\label{eqn27}
\DC_e=\lim_{N\to\infty}\frac{1}{N}\sum_{\alpha=1}^P e_{\alpha N},
\end{align}
and we only include in the sum on the right, those parties for which at least one of its local operators is not proportional to the identity. This quantity, $\DC_e$, depends on the ordering chosen. Then we have the following theorem.
\begin{thm3} \label{thm3}For any finite-round LOCC protocol of $P$ parties implementing a separable measurement corresponding to an infinite number of distinct POVM elements $\{\KC_j=\KC_j^{(1)}\otimes\ldots\otimes\KC_j^{(P)}\}$, there exists an ordering of those POVM elements such that
\begin{align}\label{eqn26}
\DC_e\le2(1-2^{-P}).
\end{align}
There exist separable measurements with an infinite number of distinct POVM elements for which the upper bound in \eqref{eqn26} can be achieved with equality.
\end{thm3}
\noindent  The proof is given in the appendix. The idea is that the LOCC protocol that implements the measurement induces an ordering for which $\DC_e$ satisfies the bound. One first prunes and rearranges the tree in a way similar to what was done for finite $N$, and then the leaves of the resulting tree can be enumerated. This enumeration provides the desired ordering. Actually, there is a great deal of freedom in choosing this enumeration, so our proof actually demonstrates that there are an infinite number of orderings for which \eqref{eqn26} is satisfied, and we can strengthen Theorem~\ref{thm3} to some degree as follows.
\begin{thm4} \label{thm4}Consider any finite-round LOCC protocol of $P$ parties implementing a separable measurement corresponding to an infinite number of distinct POVM elements $\{\KC_j=\KC_j^{(1)}\otimes\ldots\otimes\KC_j^{(P)}\}$. Then, for any finite integer $M$, and for any choice and ordering of the first $M$ of these POVM elements, there exists an ordering of the remaining POVM elements such that $\DC_e\le2(1-2^{-P})$.
\end{thm4}
\noindent The proof of this result is included in the appendix.

One can show that the bound in \eqref{eqn26} is tight by the following discussion, which closely mirrors that given above on how to achieve the upper bound with equality in the case of finite $N$. The first party makes an initial measurement with an infinite number of outcomes, each of which is followed by a sequence of $P-1$ (one for each of the other parties) two-outcome measurements along every branch. This means each outcome of that initial measurement has descended from it $2^{P-1}$ leaf nodes and a total of $2^P-1$ nodes in its descendant subtree. Choose these measurements such that all nodes are extreme rays and distinct from each other---this is not difficult to do---and then order the POVM elements in the overall separable measurement by following a right-to-left enumeration of the leaves of this LOCC tree.  Considering the subtrees descendant from the rightmost $\SC$ outcomes of the initial measurement, one has $N=2^{P-1}\SC$ leaf nodes and $(2^P-1)\SC$ extreme rays for all $P$ parties. As $\SC\to\infty$, also $N\to\infty$, and we see that $\DC_e$ of \eqref{eqn27} is equal to $(2^P-1)/2^{P-1}=2(1-2^{-P})$ for this $P$-round LOCC protocol, saturating the upper bound in \eqref{eqn26}. 

\section{Application to rank-$1$ measurements}\label{sec3}
For finite-$N$ measurements in which every operator is rank-$1$, it is a simple process to apply Theorem~\ref{thm2} to determine if these measurements are candidates for LOCC. Each rank-$1$ product operator is a product of rank-$1$ local operators, and rank-$1$ operators, being extreme rays in the full set of positive semidefinite operators, are necessarily extreme in any subset of that full set. Therefore, one need only count the number of distinct local operators in these measurements, and then violation of the bound in Theorem~\ref{thm2} automatically rules out any possibility of implementation by finite-round LOCC.

Rank-$1$ measurements arise in the context of quantum state discrimination of a full basis of any multipartite Hilbert space. When the basis is mutually orthogonal, the only\footnote{We restrict to measurements acting only on the original Hilbert space. While enlarging the Hilbert space creates the possibility of using other measurements, these other measurements are effectively identical to the ``only" measurement discussed here; see Lemma~5 of \cite{myUSD}. Therefore, enlarging the Hilbert space does not allow accomplishment by LOCC of a task that is impossible by LOCC acting only on the original Hilbert space.} measurement that can perfectly discriminate the set of states consists of rank-$1$ projectors onto the states of that basis. Clearly, these must be product states for there to be any hope of accomplishing this task by LOCC, and if they are product, Theorem~\ref{thm2} further restricts what may be possible. A well-known example of perfect discrimination of a full product basis where our results can be profitably applied is that of Bennett, et. al., which was the first demonstration of the existence of separable measurements that are not LOCC \cite{Bennett9}. This set of nine states on a $3\times3$ system, often referred to as domino states, is (omitting normalization factors)
\begin{align}\label{eqn31}
|\Psi_1\rangle & = |1\rangle|1\rangle&
|\Psi_2\rangle & = |0\rangle(|0\rangle+|1\rangle)&
|\Psi_3\rangle & = |0\rangle(|0\rangle-|1\rangle)\notag\\
|\Psi_4\rangle & = |2\rangle(|1\rangle+|2\rangle)&
|\Psi_5\rangle & = |2\rangle(|1\rangle-|2\rangle)&
|\Psi_6\rangle & = (|1\rangle+|2\rangle)|0\rangle\notag\\
|\Psi_7\rangle & = (|1\rangle-|2\rangle)|0\rangle&
|\Psi_8\rangle & = (|0\rangle+|1\rangle)|2\rangle&
|\Psi_9\rangle & = (|0\rangle-|1\rangle)|2\rangle.
\end{align}
There are seven distinct local states for each of the $P=2$ parties, so the $N=9$ separable measurement that perfectly discriminates these states involves seven distinct rank-$1$ local projectors on each side. This means that whereas $e_1=7=e_2$ and $e_1+e_2=14$, the upper bound on this sum in Theorem~\ref{thm2} is $2N-\left\lceil N/2^{P-1}\right\rceil=13$. Hence, this measurement violates Theorem~\ref{thm2}, implying directly (the well-known result) that this set of states cannot be perfectly discriminated by finite-round LOCC. The same conclusion immediately follows for any set of ``rotated domino states" \cite{ChildsLeung}, for which an arbitrary rotation is applied to each pair of superposition states (such as $|0\rangle+|1\rangle\to\cos(\theta)|0\rangle+\sin(\theta)|1\rangle,~|0\rangle-|1\rangle\to\sin(\theta)|0\rangle-\cos(\theta)|1\rangle$). Note that for the result we had obtained previously in \cite{myExtViolate1}, in which $\delta=N^{-1}$ instead of the value $\delta=2^{-P}>N^{-1}$ used here, we have a bound of $2(N-1)=16$, which does not allow a conclusion to be drawn for these states (rotated or not). Therefore, these examples demonstrate the usefulness of the extension obtained in the present paper.

\section{Conclusions}\label{sec4}
In summary, we have proved a necessary condition for any finite-round LOCC protocol, which provides an upper bound on the number of extreme rays appearing in the collection of POVM elements associated with a separable measurement, see Theorem~\ref{thm2} and the accompanying discussion. We have shown that the upper bound in Theorem~\ref{thm2} is tight for all measurements having a finite number of distinct POVM elements by providing examples of measurements for which the upper bound is achieved with equality. This has been further extended in Theorems~\ref{thm3} and \ref{thm4} to cover cases of measurements with an infinite number of distinct POVM elements, and the bound in this case can also be achieved with equality. These results extend a previous result obtained in \cite{myExtViolate1}, restated here as Theorem~\ref{thm1}, but the corresponding upper bound in that theorem is tight only when there are relatively few distinct POVM elements. 

In Section~\ref{sec2}, we have shown that the well-known separable measurement of \cite{Bennett9} violates the necessary condition of Theorem~\ref{thm2}, providing one more way of showing that this measurement cannot be implemented by finite-round LOCC. We have also noted that this measurement does not violate the condition of Theorem~\ref{thm1}, demonstrating the importance of the extension obtained in Theorem~\ref{thm2}.

We have conjectured elsewhere that Theorem~\ref{thm1} also applies to infinite-round LOCC protocols, and we continue to believe this conjecture holds. Similarly, we also believe that Theorem~\ref{thm2} holds for infinite-round protocols, but we have yet to find a proof. We feel less confident this will also be the case for Theorems~\ref{thm3} and \ref{thm4}, though it is certainly a possibility. If these conjectures turn out to be true, we will have found yet another way of proving that there is a finite gap between what can be achieved by the separable measurement which successfully distinguishes the nine states of \cite{Bennett9}, as opposed to what can be achieved by LOCC.

\noindent\textit{Acknowledgments} --- We would like to thank Li Yu and Dan Stahlke for very helpful discussions. This work has been supported in part by the National Science Foundation through Grant No. 1205931.

\appendix*\section{Proof of Theorems~\ref{thm3} and \ref{thm4}}
Our proof of Theorems~\ref{thm3} and \ref{thm4} will be very similar to that of Theorem~\ref{thm2}, except that we will not start with a canonical LOCC tree, since such a tree, being full binary, would require the infinite-leaf tree to have infinite height, whereas we wish to work with trees of finite height. According to Lemmas~$2$ and $4$ of \cite{myMany}, we may nonetheless assume that our LOCC tree is such that every nonleaf node has at least two children and that, of the POVM elements labeling its children, no two are proportional to each other.

Although the following lemma applies only to trees with a finite number of leaf nodes, it will play an important role in our arguments.
\begin{lem1}\label{lem1}
For any tree of height $h$ in which every nonleaf node has at least two children, the ratio of the total number of nodes $n$ to the number of leaf nodes $l$ in the tree satisfies
\begin{align}\label{eqnA1}
\frac{n}{l}\le2\left(1-2^{-(h+1)}\right),
\end{align}
as long as $l$, and hence $n$, is finite.
\end{lem1}
\proof The proof is by induction on the height $h$. For $h=1$, the tree has a root node and $l\ge2$ leaf nodes, for a total of $n=l+1$ nodes in all. Then, $n/l=1+1/l\le3/2=2\left(1-2^{-\left(h+1\right)}\right)$. Now assume \eqref{eqnA1} holds for $h=H-1$, and let us show that it then holds for $h=H$. Let $T_H$ be a tree of height $H$ obtained from $T_{H-1}$ by adding children to some of the leaf nodes of $T_{H-1}$. Those leaf nodes to which we do not add children are terminal at $H-1$, let there be $t_{H-1}$ of these terminal leafs. If we add the number of leaf nodes from $T_H$ to the total number of nodes in $T_{H-1}$, we overcount the total number of nodes in $T_H$ because those terminal leafs have been counted twice. Therefore, the total number of nodes in $T_H$ is $n_H=n_{H-1}+l_H-t_{H-1}$. In addition, since we consider only trees for which each nonleaf node has at least two children, we have $l_H\ge2\left(l_{H-1}-t_{H-1}\right)+t_{H-1}=2l_{H-1}-t_{H-1}$. Hence, defining $x=t_{H-1}/l_{H-1}$, we have
\begin{align}\label{eqnA3}
\frac{n_{H}}{l_{H}}&=1+\frac{n_{H-1}-t_{H-1}}{l_{H}}\le1+\frac{n_{H-1}-t_{H-1}}{2l_{H-1}-t_{H-1}}=1+\frac{n_{H-1}/l_{H-1}-x}{2-x}\notag\\
										&\le1+\frac{2\left(1-2^{-H}\right)-x}{2-x}=2-\frac{2^{-(H-1)}}{2-x}\le2-2^{-H}=2\left(1-2^{-(H+1)}\right),
\end{align}
where the first inequality on the second line is by the induction assumption, and this completes the proof.\endproof

\noindent\textit{Proof of Theorem~\ref{thm3}.} Consider any finite-round LOCC tree implementing a separable measurement defined by the infinite set of POVM elements $\{\KC_j\}$. This tree has an infinite number of leaf nodes, at least one for each $\KC_j$. We prune this tree following the technique of \cite{myExtViolate1}, except that if at any stage of this process we are removing a subtree whose root has more than one sibling, then we simply remove that subtree without removing an additional nonleaf node (since in \cite{myExtViolate1} the tree was full binary, the subtrees considered there always had one and only one sibling; it was then necessary to remove an extra nonleaf node in order to keep the tree full binary; see \cite{myExtViolate1} for details). If that subtree has only one sibling, then remove it according to the rules used in \cite{myExtViolate1}. The pruning is complete when there is one and only one leaf for each of the $\KC_j$. According to this procedure, every nonleaf node in the resulting tree still has at least two children.

The next step is to rearrange the resulting tree in the same way we did for the finite-$N$ case, exchanging an extreme node with one of its non-extreme children, if there is one, and continuing this process until no extreme node has a non-extreme descendant. The tree that remains has all its extreme nodes in subtrees within which every node is extreme, and just as in the finite-$N$ case, these subtrees can have height no greater than $P-1$.

Choose any one of these subtrees and set $\SC=1$. If this is a finite subtree we can include it in its entirety from the outset, so add another subtree to the collection and increment $\SC$. If instead it is an infinite subtree, we will need to count its nodes using some kind of a limiting procedure. Hence for each infinite subtree, instead of starting with the entire subtree, add it in as a ``skeleton" of itself, one which is a full binary tree. Any such skeleton will do, as long as every branch in it is also a branch in the original subtree. These skeletons may be obtained from their corresponding subtree by removing all but two children from every node that has more than two, while also removing the complete branches descended from those removed children. At each subsequent step, include another subtree in the collection and increment $\SC$. At the same time, for each skeleton of an infinite subtree $T_s$, add a full binary branch to that skeleton, by which we mean a branch for which every nonleaf node has two children, where the added branch is either one that was in the original $T_s$, or a skeleton of one that was. Add these skeletal branches in the order indicated by index $s$, starting at the infinite subtree with smallest $s$ and proceeding to the one with next smallest $s$, and so on. Continue this process of adding subtrees and branches indefinitely. In the limit of an infinite number of steps of this procedure, each $T_s$ will be fully reconstructed and every subtree will be included in the collection. If all subtrees are finite, there will be an infinite number of subtrees to include, one at each step. Otherwise, there may be a finite or infinite number of subtrees to include, but reconstruction of the infinite subtrees will always require an infinite number of steps. In any case, at each step of this infinitely long process, we have a finite number $\SC$ of subtrees, each having $l_s$ leaf nodes and $n_s$ nodes in total, with both $l_s$ and $n_s$ finite. 

We need to identify a precise ordering of the $\KC_j$. Such an ordering may be obtained directly from the procedure described above of including more and more subtrees, while at the same time reconstructing each infinite subtree in a step-by-step fashion. In fact, this procedure generates an infinite number of such orderings. The index $s$, which can be assigned arbitrarily, provides a kind of coarse-grained order for the $\KC_j$, indicating when each finite subtree is added, when each infinite one is begun as a skeleton, and also the order in which each additional skeletal branch is added to those infinite subtrees previously begun. There still remains the task of ordering the set of $\KC_j$ within each of these ``coarse-grained" objects. Note that each of the $\KC_j$ appears in one and only one of the subtrees (recall that the pruned tree has one and only one appearance of each $\KC_j$), so this fine-grained ordering will be unambiguous. For each skeletal branch then, choose any ordering that has the $\KC_j$ that appear within it ordered one right after another, which then ensures that there is no more than one branch at a time in the entire collection of (partially reconstructed) subtrees that is not full binary. This means that at each step of this procedure, every nonleaf node in the entire collection has at least two children, except those nodes in the branch that is presently being constructed.\footnote{To be more precise about this, for each new subtree, start with any one leaf that was at the end of a branch of height $h\le P-1$ the same as that of the original subtree, adding this solitary leaf along with its $h-1$ ancestors, one of which is the root of that subtree. The next leaf is chosen as one whose branch attaches to that preceding branch (which will add no more than $h-1$ nodes to this subtree, including that leaf, since it must share at least one node with the preceding branch to which it attaches). Subsequent leafs are chosen to attach to this same skeleton in a way such that no node in it has more than two children, and this continues until every nonleaf node has two. Then, move on to the next subtree. If a subtree has already been started, then it has a full binary skeleton already present, so add any additional leaf to start the next skeleton. This leaf attaches to that full binary skeleton at a node that already had at least two children, so will now have more than two, but in general, this new leaf will have ancestors that have only one child node. Continue adding leafs to the skeleton consisting of that leaf and its ancestors until it is also full binary, and then move on to the next subtree, continuing this process indefinitely.}

At any given point, let $s_\ast$ denote the one subtree that has a branch that is not yet partially completed to full binary. Let $\delta n$ be the number of nodes on the skeletal branch in this subtree that is presently being constructed and is not yet part of a full binary skeleton, and let $\delta l$ be the corresponding number of leafs. Given that these branches have height no greater than $P-1$, then the number of leaf nodes in the skeletal branch that is not yet full binary must satisfy $\delta l\le 2^{P-1}$. Define $n_C=n_{s_\ast}-\delta n$ and $l_C=l_{s_\ast}-\delta l$. Then since $n_C,l_C$ count the nodes and leafs that lie in branches for which every nonleaf node has at least two children, we have from Lemma~\ref{lem1} that $n_C/l_C\le2\left(1-2^{P-1}\right)$. Now, each time one adds a leaf, one adds no more than $P$ nodes, strictly fewer than this if that leaf is attaching to a subtree already begun. Therefore, $\delta n/\delta l\le P$. 

Define $N=\sum_sl_s$, which is the number of distinct $\KC_j$ appearing in the collection of subtrees at this stage of the process. The total number of extreme rays appearing in this collection is no greater than the total number of nodes, $\sum_\alpha e_{\alpha N}\le\sum_sn_s$. Then, for any ordering as described above, we have
\begin{align}\label{eqnA4}
\frac{1}{N}\sum_{\alpha}e_{\alpha N}&\le\sum_{s=1}^\SC n_s\bigg/\sum_{s=1}^\SC l_s\notag\\
			&=\sum_{s\neq s_\ast}^\SC l_s \left(\frac{n_s}{l_s}\right)\bigg/\sum_{s=1}^\SC l_s +l_C\left(\frac{n_C}{l_C}\right)\bigg/\sum_{s=1}^\SC l_s+\delta l\left(\frac{\delta n}{\delta l}\right)\bigg/\sum_{s=1}^\SC l_s \notag\\		
			&\le2\left(1-2^{-P}\right)\left(\sum_{s\neq s_\ast}^\SC l_s+l_C\right)\bigg/\sum_{s=1}^\SC l_s +P\delta l\bigg/\sum_{s=1}^\SC l_s \notag\\
			&=2\left(1-2^{-P}\right)+\left(P-2+2^{-(P-1)}\right)\delta l\bigg/\sum_{s=1}^\SC l_s\notag\\
			&\le2\left(1-2^{-P}\right)+\left(P-2+2^{-(P-1)}\right)2^{P-1}\bigg/\sum_{s=1}^\SC l_s.
\end{align}
where the third line follows from Lemma~\ref{lem1}, which tells us that $n_s/l_s\le2\left(1-2^{-P}\right)$ for all $s\ne s_\ast$ and that $n_C/l_C\le2\left(1-2^{-P}\right)$, along with the fact that $\delta n/\delta l\le P$, as argued above. The last line follows from $\delta l\le2^{P-1}$, which was also argued above. Now as $\SC\to\infty$,  $N=\sum_{s=1}^\SC l_s\to\infty$. Hence in this limit, we see that the second term in the last line approaches zero, and we recover $\DC_e\le2\left(1-2^{-P}\right)$. This completes the proof of Theorem~\ref{thm3}.\endproof

\noindent\textit{Proof of Theorem~\ref{thm4}.} Theorem~\ref{thm4} follows almost immediately from the proof of Theorem~\ref{thm3}. For any ordering of the first $M$ of the $\KC_j$, fill in the subtrees constructed from those $M$ leafs until they are full binary. Then, for the remaining leafs, continue precisely as described in the proof of Theorem~\ref{thm3}. The result follows directly.~\endproof


\end{document}